\documentclass[aps,twocolumn,groupedaddress]{revtex4}

\usepackage{graphicx}
\usepackage{dcolumn}
\usepackage{bm}
\usepackage{amsmath}

\begin{document}


\title{Intrinsic and Rashba Spin-orbit Interactions in Graphene Sheets}
\author{Hongki Min}
\email{hongki@physics.utexas.edu}
\affiliation{Department of Physics, The University of Texas at Austin, Austin Texas 78712}
\author{J.E. Hill}
\affiliation{Department of Physics, The University of Texas at Austin, Austin Texas 78712}
\author{N.A. Sinitsyn}
\affiliation{Department of Physics, The University of Texas at Austin, Austin Texas 78712}
\author{B.R. Sahu}
\affiliation{Department of Physics, The University of Texas at Austin, Austin Texas 78712}
\author{Leonard Kleinman}
\affiliation{Department of Physics, The University of Texas at Austin, Austin Texas 78712}
\author{A.H. MacDonald}
\affiliation{Department of Physics, The University of Texas at Austin, Austin Texas 78712}
\date{June 20, 2006}

\begin{abstract}
Starting from a microscopic
tight-binding model and using second order perturbation theory, we derive explicit expressions for the
intrinsic and Rashba spin-orbit interaction induced gaps
in the Dirac-like low-energy band structure of an isolated graphene sheet.
The Rashba interaction parameter is first order
in the atomic carbon spin-orbit coupling strength $\xi$ and first order in the
external electric field $E$ perpendicular to the graphene plane, whereas the intrinsic spin-orbit interaction
which survives at $E=0$ is second order in $\xi$.  The spin-orbit terms in the low-energy effective
Hamiltonian have the form proposed recently by Kane and Mele.
\textit{Ab initio} electronic structure calculations were performed as a partial
check on the validity of the tight-binding model.
\end{abstract}

\maketitle

\section{Introduction}
Graphene is a two-dimensional honeycomb lattice of carbon atoms that has attracted considerable attention
recently because of experimental progress\cite{novoselov2004,novoselov2005a,zhang2005a,berger2004} that has raised hopes
for applications in nanoelectronics
and because of exotic chiral features\cite{kane2005a,kane2005b,semonoff1984,haldane1988,gusynin2005,sinitsyn2006,saito1998,wilson2006} in its electronic structure.
In the absence of spin-orbit interactions, the energy bands of graphene are described at low energies by a
two-dimensional Dirac equation with linear dispersion centered on the hexagonal corners of the
honeycomb lattice Brillouin zone.
The recent advances in fabrication techniques have made it possible to produce graphitic systems with only a few layers
or even a single monolayer of graphene\cite{novoselov2004,novoselov2005a,zhang2005a,berger2004}.

One of the most remarkable properties of graphene is its half integer
quantum Hall effect, confirmed by recent experiments\cite{novoselov2005b,zhang2005b}.
This electronic property follows directly from the system's Dirac-like band structure\cite{kane2005a,kane2005b}.
In a recent paper, Kane and Mele\cite{kane2005a} showed that symmetry allowed spin-orbit interactions can generate an energy gap
and convert graphene from a two dimensional zero gap semiconductor to an insulator with a quantized spin Hall effect\cite{sinitsyn2006}.
The quantized spin Hall conductivity can be zero or nonzero, depending on the relative strength of intrinsic
and Rashba spin-orbit interactions.  The temperature at which the spin Hall effect
can be observed, and the sample quality requirements for its occurrence, depend
on the absolute magnitude of these two spin-orbit interaction terms in the band structure.  (Kane and Mele\cite{kane2005a}
argued on the basis of rough estimates of the spin-orbit interaction scale, that the quantum spin Hall effect in graphene
should be observable at relatively accessible temperatures of the order of 1$^\circ {\rm K}$.)
Motivated by the fundamental interest associated with the spin Hall effect and spin-orbit interactions in graphene,
we have attempted to estimate, on the basis of microscopic considerations,
the strength of both interactions.

In order to allow for a Rashba interaction, we account for the presence of an external gate
electric field $E$ of the type used experimentally in graphene to move the Fermi energy away from the Dirac point.  (Importantly
this electric field explicitly removes inversion through the graphene plane from the symmetry operations of the system.)
Then, starting from a microscopic tight-binding model with atomic spin-orbit interactions of strength $\xi$, we use perturbation theory to
derive expressions for the spin-orbit coupling terms that appear in the low-energy Hamiltonian.  At leading order in
$\xi$ only the Rashba spin-orbit interaction term ($\propto E$) appears.  The intrinsic ($E=0$) spin-orbit
coupling has a leading contribution proportional to $\xi^2$.  Both terms have the form proposed by Kane and
Mele\cite{kane2005a} on the basis of symmetry considerations.  According to our theory the respective coupling constants
are given by the following expressions:
\begin{equation}
\label{eq:intrinsic}
\lambda_{SO} = {|s| \over 18(sp\sigma)^2} \; \xi^2 \ , \
\end{equation}
and
\begin{equation}
\label{eq:rashba}
\lambda_R = { e E z_0 \over 3(sp\sigma)} \; \xi,
\end{equation}
where $|s|$ and $(sp\sigma)$ are tight-binding model parameters explained more fully below, $E$ is
a perpendicular external electric field, and $z_0$ is
proportional to the atomic size of carbon.
The coupling constants $\lambda_{SO}$ and $\lambda_R$ have numerical values $\sim 100$ times smaller
and $\sim 100$ times larger, respectively, than the estimates of Kane and Mele\cite{kane2005a} with $\lambda_{SO}<\lambda_R$
at the largest reasonable values of $E$.  Together,
these estimates suggest that the quantum spin Hall effect will be observable
in ideal samples only at temperatures below $~\sim 0.01 ^\circ {\rm K}$ in a zero-field limit.

Our paper is organized as follows.  In Section \ref{sec:tight_binding} we briefly summarize the tight-binding model used to
represent graphene in this paper.  Section \ref{sec:perturbation} describes some details of the perturbation theory calculation.
In Section \ref{sec:abinitio} we discuss \textit{ab initio}
density functional theory calculations we have carried out as a partial check on the tight-binding model and on the atomic approximation for spin-orbit
interactions used in the perturbation theory calculations.  We conclude in Section \ref{sec:summary} with a brief summary and
present our conclusions.

\section{Tight-binding Model}
\label{sec:tight_binding}
\subsection{Two-center hopping}
For our analytic perturbation theory calculations we choose the simplest
possible tight-binding model with carbon $s$ and $p$
orbitals, a two-center Slater-Koster approximation\cite{slater1954} for nearest-neighbor hopping, and orthogonality between Wannier functions centered
on different sites assumed.  This gives a tight-binding Hamiltonian of the form
\begin{eqnarray}
\label{eq:NN_matrix}
H_{A,\mu;A,\mu'}(\vec{k})=H_{B,\mu;B,\mu'}(\vec{k})&=&t_{\mu}\delta_{\mu,\mu'}, \\
H_{A,\mu;B,\mu'}(\vec{k})=H_{B,\mu';A,\mu}^{\ast}(\vec{k})&=&\sum_{i=1}^{3} e^{i\vec{k}\cdot\vec{N}_i} t_{\mu,\mu'}(\vec{N}_i), \nonumber
\end{eqnarray}
where $\mu,\mu'$ label the four orbitals on each site,
$A$ and $B$ represent the two distinct sites in the honeycomb lattice unit cell, and $\vec{N}_i$ is one of the
three vectors connecting a lattice site and its near neighbors.  We choose a coordinate system in which the honeycomb's
Bravais lattice has primitive vectors
\begin{equation}
\vec{a}_1=a(1,0) \ , \qquad\qquad \vec{a}_2=a\Big({1 \over 2},{\sqrt{3} \over 2}\Big),
\end{equation}
where $a = 2.46 {\rm \AA}$ is the lattice constant of graphene.  The corresponding reciprocal lattice vectors are
\begin{equation}
\vec{b}_1={4\pi \over \sqrt{3}a}\Big({\sqrt{3}\over 2},-{1\over 2}\Big) \ , \, \vec{b}_2={4\pi \over \sqrt{3}a}(0,1),
\end{equation}
and the near-neighbor translation vectors are:
\begin{equation}
\label{eq:NN}
\vec{N}= \left\{a\Big(0,{1 \over\sqrt{3}}\Big),a\Big(-{1\over 2},-{1\over 2\sqrt{3}}\Big),a\Big({1\over 2},-{1\over 2\sqrt{3}}\Big) \right\}.
\end{equation}
The site-diagonal matrix elements $t_{\mu}$ are the atomic energies of $s$ and $p$ orbitals, with the latter chosen as the
zero of energy.  In Table \ref{tab:two_center} we reproduce for completeness the relationship between the required nearest-neighbor hopping
matrix elements $t_{\mu,\mu'}$ and the four independent Slater-Koster parameters $(ss\sigma)$, $(sp\sigma)$, $(pp\sigma)$, and
$(pp\pi)$ whose numerical values specify this model quantitatively.
If the graphene lattice is placed in the $\hat{x}-\hat{y}$ plane, $n_z=0$ for hops on the graphene lattice
and the atomic $p_z$ orbitals decouple from other orbitals.
This property is more general than our model, since it follows from the graphene plane inversion symmetry that
orbitals which are even and odd under this symmetry operation will not be coupled, and is
key to the way in which weak spin-orbit interactions influence the low-energy bands.

\begin{table}[h]
\begin{ruledtabular}
\begin{tabular}{cc|cc}
$t_{s}$     & $s$              & $t_{p_x,p_x}$ & $n_x^2(pp\sigma)+(1-n_x^2)(pp\pi)$   \\
$t_{p}$     & $p$              & $t_{p_y,p_y}$ & $n_y^2(pp\sigma)+(1-n_y^2)(pp\pi)$   \\
$t_{s,s}  $ & $(ss\sigma)$     & $t_{p_z,p_z}$ & $n_z^2(pp\sigma)+(1-n_z^2)(pp\pi)$   \\
$t_{s,p_x}$ & $n_x (sp\sigma)$ & $t_{p_x,p_y}$ & $n_x n_y (pp\sigma)-n_x n_y (pp\pi)$ \\
$t_{s,p_y}$ & $n_y (sp\sigma)$ & $t_{p_x,p_z}$ & $n_x n_z (pp\sigma)-n_x n_z (pp\pi)$ \\
$t_{s,p_z}$ & $n_z (sp\sigma)$ & $t_{p_y,p_z}$ & $n_y n_z (pp\sigma)-n_y n_z (pp\pi)$ \\
\end{tabular}
\caption{Two-center matrix elements for hoping between $s$ and $p$ orbitals along
a direction specified by the unit vector $(n_x,n_y,n_z)$.}
\label{tab:two_center}
\end{ruledtabular}
\end{table}

\subsection{Atomic spin-orbit interactions}
The microscopic spin-orbit interaction is
\begin{equation}
H_{SO}={1 \over 2 (m_e c)^2}\left(\nabla V\times\vec{p}\right)\cdot\vec{S}.
\end{equation}
Since $\nabla V$ is largest near the atomic nuclei, spin-orbit interactions
are normally accurately approximated by a local atomic contribution of the form:
\begin{equation}
H_{SO}= \sum_{i,l} \; P_{il} \; \xi_l \; \vec{L}_{i}\cdot\vec{S}_{i},
\end{equation}
where $i$ is a site index, $P_{il}$ denotes projection onto angular momentum $l$ on site $i$, $\xi_l$ is the
atomic spin-orbit coupling constant for angular momentum $l$, and $\vec{S}$ is the spin operator on site $i$.
For our model spin-orbit coupling occurs only among the $p$ orbitals.

\subsection{External gate electric fields}
Finite carrier densities have been generated in graphene by applying an external gate voltage.  The resulting electric field $E$ lifts
inversion symmetry in the graphene plane.  An electric field $E$ can also be produced by accidental doping in the substrate or cap
layer or by atomic length scale charge rearrangements near the graphene/substrate or graphene/cap-layer interfaces.  To
model this important effect we consider an additional local atomic single-particle Stark-effect term of the form
\begin{equation}
H_{EF}=e E \sum_i z_i
\end{equation}
where $i$ is a site index.  In our $s-p$ tight-binding model
the only nonvanishing matrix element of $H_{EF}$ is the one between the $s$ and $p_{z}$ orbitals
to which we assign the value $e E z_0$.

\section{Perturbation Theory}
\label{sec:perturbation}
\subsection{Unperturbed Hamiltonian matrix at $K$ and $K'$}
The low-energy Hamiltonian is specified by the Dirac Hamiltonian and by the spin-orbit coupling terms at
$K$ and $K'$.  We choose the inequivalent hexagonal corner wave vectors $K$ and $K'$
to be $K={1\over 3}(2\vec{b}_1+\vec{b}_2)=({4\pi\over 3a},0)$ and $K'=-K$.
Table \ref{tab:hamiltonian_K} and Table \ref{tab:eigenvector_K} list the Hamiltonian matrix elements and
the corresponding eigenvectors. Here $s$ is the on-site energy of $s$ orbitals relative to $p$ orbitals, $\alpha\equiv {3\over 2}(sp\sigma)$, $\beta\equiv{3\over 4}\left[(pp\sigma)-(pp\pi)\right]$, and $\gamma_{\pm}={\sqrt{s^2+8\alpha^2}\pm s \over 2}$.
\begin{table}[h]
\begin{ruledtabular}
\begin{tabular}{c|cccc|cccc}
Orbital     &$A,s$ & $A,p_x$ & $A,p_y$ & $A,p_z$ & $B,s$   & $B,p_x$ & $B,p_y$ & $B,p_z$  \\
\hline
$A,s$   &s &0 &0 &0  &0 &$\pm i\alpha$ &$\alpha$ &0\\
$A,p_x$ &0 &0 &0 &0  &$\mp i\alpha$ &$-\beta$ &$\mp i\beta$ &0 \\
$A,p_y$ &0 &0 &0 &0  &$-\alpha$ &$\mp i\beta$ & $\beta$ &0 \\
$A,p_z$ &0 &0 &0 &0  &0 &0 &0 &0 \\
\hline
$B,s$   &0 &$\pm i\alpha$ & $-\alpha$ &0           &s &0 &0 &0 \\
$B,p_x$ &$\mp i\alpha$ &$-\beta$ &$\pm i\beta$ &0  &0 &0 &0 &0 \\
$B,p_y$ &$\alpha$ &$\pm i\beta$ &$\beta$ &0        &0 &0 &0 &0 \\
$B,p_z$ &0 &0 &0 &0                                &0 &0 &0 &0 \\
\end{tabular}
\caption{Tight-binding model matrix elements at the $K$ and $K'$ points in the absence of
spin-orbit interactions and external electric fields.  The first (second) sign corresponds to the $K(K')$ point.}
\label{tab:hamiltonian_K}
\end{ruledtabular}
\end{table}
Note that the $\sigma$ bands are decoupled from the $\pi$ bands.
When the spin-degree of freedom is included, the $E=0$ eigenstates at $K$ and $K'$ are fourfold degenerate.
Below we refer to this degenerate manifold as $D$.
\begin{table}[h]
\begin{ruledtabular}
\begin{tabular}{c|cccccccc}
E	        &$A,s$          &$A,p_x$        &$A,p_y$        &$A,p_z$        &$B,s$          &$B,p_x$        &$B,p_y$        &$B,p_z$        \\
\hline
$-\gamma_{-}$   &$-\gamma_{-}$  &0              &0              &0              &0              &$\mp i\alpha$  &$\alpha$       &0              \\
$-\gamma_{-}$   &0              &$\mp i\alpha$  &$-\alpha$      &0              &$-\gamma_{-}$  &0              &0              &0              \\
$-2\beta$       &0              &$\pm i$        &$-1$           &0              &0              &$\pm i$        &1              &0              \\
0               &0              &0              &0              &1              &0              &0              &0              &0              \\
0               &0              &0              &0              &0              &0              &0              &0              &1              \\
$\gamma_{+}$    &$\gamma_{+}$   &0              &0              &0              &0              &$\mp i\alpha$	&$\alpha$	&0              \\
$\gamma_{+}$    &0              &$\mp i\alpha$  &$-\alpha$      &0              &$\gamma_{+}$   &0              &0              &0              \\
$2\beta$        &0              &$\mp i$        &1              &0              &0              &$\pm i$        &1              &0              \\
\end{tabular}
\caption{Unnormalized unperturbed eigenvectors at the $K$ and $K'$ points arranged in increasing order of energies
assuming $0<\gamma_{+}<2\beta<\gamma_{-}$. The first (second) sign corresponds to the $K$($K'$) point.}
\label{tab:eigenvector_K}
\end{ruledtabular}
\end{table}

%
%

\subsection{Low-energy effective Hamiltonian}
We treat the atomic spin-orbit interaction and the external electric fields as a perturbation:
\begin{equation}
\Delta H= H_{SO}+H_{EF}.
\label{dH}
\end{equation}
The effective Hamiltonian which lifts the $E=0$ degeneracy is given by the second-order degenerate state perturbation theory expression\cite{schiff1968}:
\begin{equation}
\label{eq:H_second}
H_{m,n}^{(2)}=\sum_{l\not\in D} {\left<m^{(0)}\right|\Delta H\left|l^{(0)}\right> \left<l^{(0)}\right|\Delta H\left|n^{(0)}\right> \over E_D - E_l^{(0)} }
\end{equation}
where $m,n\in D$.  An elementary calculation then shows that the matrix elements of $H_{m,n}^{(2)}$ (at the $K$ point) are
those listed in Table \ref{tab:effective_K} with $\lambda_{SO}$ and $\lambda_R$ defined by Eqs. (\ref{eq:intrinsic}) and (\ref{eq:rashba}) respectively.

\begin{table}[h]
\begin{ruledtabular}
\begin{tabular}{c|c|c|c|c}
Orb                 &   $A,p_{z,\uparrow}$& $A,p_{z,\downarrow}$&   $B,p_{z,\uparrow}$& $B,p_{z,\downarrow}$ \\
\hline
$A,p_{z,\uparrow}$  &0              &0               &0               &0               \\
$A,p_{z,\downarrow}$&0              &$-2\lambda_{SO}$&$2i\lambda_R$   &0               \\
$B,p_{z,\uparrow}$  &0              &$-2i\lambda_R$  &$-2\lambda_{SO}$&0               \\
$B,p_{z,\downarrow}$&0              &0               &0               &0               \\
\end{tabular}
\caption{The effective spin-orbit matrix at the $K$ point.}
\label{tab:effective_K}
\end{ruledtabular}
\end{table}


Similar results are obtained at the $K'$ point.  It follows that
the effective spin-orbit interaction for $\pi$ orbitals is
\begin{equation}
\label{eq:effective}
H_{eff}= -\lambda_{SO} + \lambda_{SO} \;  \sigma_z \tau_z s_z + \lambda_R \; \big(\sigma_x \tau_z s_y - \sigma_y s_x \big)
\end{equation}
where the $\sigma_{\alpha}$ Pauli matrices act in the $A,B$ space with $\sigma_{z}$ eigenstates localized on a
definite site, $\tau_z = \pm 1$ for $K,K'$ points, and 
the $s_{\alpha}$ are Pauli matrices acting on the electron's spin.
This Hamiltonian differs from the form proposed by Kane and Mele\cite{kane2005a} only by the constant
$-\lambda_{SO}$.  The excitation spectrum has a gap $E_{gap} = 2 (\lambda_{SO}-\lambda_{R})$
and the system has a quantized spin Hall effect\cite{kane2005a} for $0< \lambda_{R} < \lambda_{SO}$.

To obtain quantitative estimates for the coupling constants we used the tight-binding
model parameters listed in Table \ref{tab:hopping},
taken from Ref. \onlinecite{saito1992}. For the spin-orbit coupling parameter among the $p$ orbitals we use $\xi=6$ meV,
a value obtained by fitting carbon atomic energy levels given by the \textit{ab initio} electronic structure code
described below.
\begin{table}[h]
\begin{ruledtabular}
\begin{tabular}{c|c|c}
Parameter	&Energy(eV)	&Overlap	\\
\hline
$s$		&-8.868		&1		\\
$p$		&0		&1		\\
$ss\sigma$	&-6.769		&+0.212		\\
$sp\sigma$	&+5.580		&-0.102		\\
$pp\sigma$	&+5.037		&-0.146		\\
$pp\pi$		&-3.033		&+0.129		\\
\end{tabular}
\caption{Hopping parameters for a graphene taken from Ref. \onlinecite{saito1992}.}
\label{tab:hopping}
\end{ruledtabular}
\end{table}
These values imply a graphene energy gap at $\lambda_{R}=0$ equal to
\begin{equation}
\label{eq:gap_estimate}
2\lambda_{SO}={|s| \over 9(sp\sigma)^2}\xi^2 \approx 0.00114 \;{\rm meV} \approx k_B \times 0.0132 ^\circ {\rm K},
\end{equation}
where we used the nonorthogonal tight-binding parameters neglecting the overlap for simple estimations.
Our estimates of $\lambda_R$ are discussed later.

\section{Ab initio Calculations}
\label{sec:abinitio}
We have performed realistic \textit{ab initio} electronic structure calculations\cite{kohn1965}
for inversion symmetric ($\lambda_R = 0$) graphene sheets using the projector augmented wave (PAW)\cite{blochl1994}
method with a Perdew-Burke-Ernzerhof (PBE) generalized gradient approximation (GGA)\cite{perdew1996} density functional
in order to partly test the quantitative accuracy of the conclusions reached here about spin-orbit interaction
gaps based on a simplified electronic structure model.  The calculations were performed using VASP
(Vienna \textit{ab initio} simulation package)\cite{kresse1996}.
In VASP, spin-orbit interactions are implemented in the PAW method which is based on a transformation
that maps all electron wave functions to smooth pseudo wave functions. All
physical properties are evaluated using pseudo wave functions.
The spin-orbit interaction is evaluated taking into account only the spherical part of the potential inside
muffin tins surrounding the carbon nuclei:
\begin{equation}
H_{SO}={1 \over 2 (m_e c)^2}{1\over r}{dV \over dr} \vec{L}\cdot\vec{S}.
\end{equation}
In order to make the gaps induced by spin-orbit interaction exceed the accuracy of VASP eigenvalues,
we have artificially increased the strength of $H_{SO}$ by up to 300 times by decreasing the
speed of light $c$.

\begin{figure}[h]
\scalebox{0.45}{\includegraphics{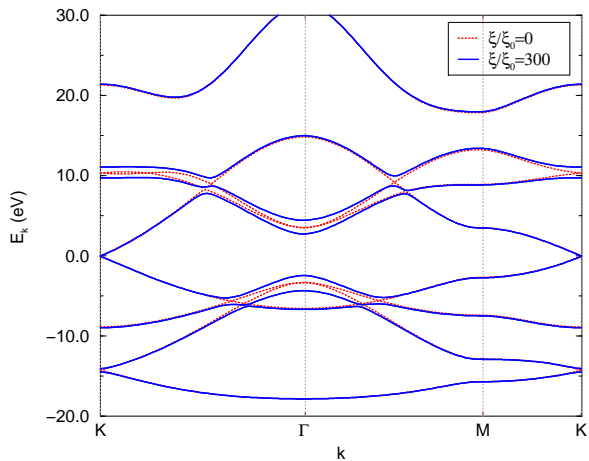}}
\caption{(Color online) Graphene band structure for $\xi=0$ and $\xi=300\xi_0$ using the tight-binding model with nonorthogonal orbitals. Hopping parameters were taken from Ref. \onlinecite{saito1992} and $\xi_0=6$ meV was used for the atomic spin-orbit coupling strength.}
\label{fig:band_SO}
\end{figure}

Figure \ref{fig:band_SO} shows the tight-binding band structure of graphene for $\xi=0$ and $\xi=300\xi_0$, where $\xi_0=6$ meV.
The spin-orbit gap is not large on the scale of the full band width, even when enlarged by a factor of 300.

\begin{figure}[h]
\scalebox{0.45}{\includegraphics{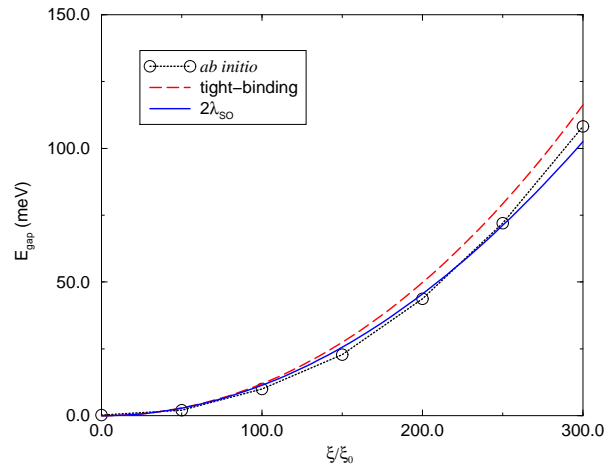}}
\caption{(Color online) Energy gap for $\lambda_R=0$ as a function of spin-orbit coupling strength from the \textit{ab initio} calculation,
from the tight-binding model with nonorthogonal orbitals, and from the analytic expression in Eq.(\ref{eq:intrinsic}).}
\label{fig:gap_SO}
\end{figure}

Figure \ref{fig:gap_SO} compares the \textit{ab initio} calculation and tight-binding model low-energy gaps at the hexagonal Brillouin-zone
corners for $\lambda_{R}=0$, finding close agreement.  Both approximations find a gap that grows as
the second power of the spin-orbit coupling strength.  The close agreement is perhaps not surprising given that VASP also
makes an atomiclike approximation for the spin-orbit coupling strength.  In our opinion, however, the neglected
contributions from interstitial regions and from aspherical potentials inside the muffin-tin sphere are small
and their contributions to energy levels tends toward even smaller values due to spatial averaging by the
Bloch wave functions.  We believe that these calculations demonstrate that the tight-binding model spin-orbit
gap estimates are accurate.

\section{Discussion and Summary}
\label{sec:summary}

The intrinsic and Rashba spin-orbit interactions arise from mixing between $\pi$ and $\sigma$ bands due to
atomic spin-orbit interactions alone in the case of $\lambda_{SO}$ (Eq.(\ref{eq:intrinsic})) and due to
a combination of atomic spin-orbit and Stark interactions in the case of $\lambda_{R}$ (Eq.(\ref{eq:rashba})).
These expressions for $\lambda_{SO}$ and $\lambda_{R}$ follow directly from Eq.(\ref{eq:H_second}) and from the
eigenvectors and eigenenergies listed in Table \ref{tab:eigenvector_K}.  (The energetic ordering in
Table \ref{tab:eigenvector_K} applies for $0<\gamma_{+}< 2\beta< \gamma_{-}$ which holds for the
tight-binding parameters in Table \ref{tab:hopping}.)
The pure $p-p$ hybridized bonding and antibonding states (energies $\pm 2 \beta$ in Table \ref{tab:eigenvector_K})
are symmetrically spaced with respect to the undoped Fermi level and do not make a net contribution
to either $\lambda_{R}$ or $\lambda_{SO}$.  The $s-p$ hybridized bonding states (energy
$- \gamma_{-}$ in Table \ref{tab:eigenvector_K}), on the other hand, are
further from the Fermi energy than the corresponding antibonding states (energy $+\gamma_{+}$ in Table \ref{tab:eigenvector_K}) because
of the difference between atomic $s$ and $p$ energies.  Their net contribution to $\lambda_{SO}$
is proportional to $s$ and inversely related to $(sp\sigma)$, which sets the scale of the
energy denominators.  Similar considerations explain the expression for $\lambda_{R}$ which is
proportional to $\xi$ and $eEz_0$ and inversely proportional to $sp\sigma$.
$\lambda_{R}$ vanishes at $E=0$ because of the inversion symmetry of an isolated graphene plane.

The numerical value of the Rashba interaction parameter $\lambda_R$ obviously depends on the electric field perpendicular to the
graphene plane which varies as the carrier density is modulated by a gate voltage.  A typical value
can be crudely estimated from Eq. (\ref{eq:rashba}), by assuming a typical electric field $E\sim \;{\rm 50V/300nm}$\cite{kane2005a},
and using the value $z_0\sim 3a_B\times \big(0.620 \AA /0.529 \AA\big)$ obtained by scaling the
hydrogenic orbital Stark matrix element by the ratio of the atomic radii\cite{waber1965} of carbon and hydrogen:
\begin{equation}
\lambda_R={ e E z_0 \over 3(sp\sigma)} \; \xi \approx 0.0111 \;{\rm meV} \approx k_B \times 0.129 ^\circ {\rm K}.
\end{equation}
The value of $\lambda_R$ is influenced by screening of the electric field at one graphene atom by
the polarization of other graphene atoms and by dielectric screening in the substrate and cap layers,
but these correction factors are expected to be $\sim 1$.
Note that our estimate for $\lambda_{SO}$ is 100 times smaller than Kane's estimate,$\sim 1^\circ {\rm K}$, whereas
$\lambda_R$ is 100 times larger than Kane's estimate, $\sim 0.5^\circ {\rm mK}$.  If our estimates
are accurate, $\lambda_{SO}<\lambda_R$ at large gate voltages.  For undoped samples, however,
the requirement for a quantized spin Hall effect gap\cite{kane2005a},
$\lambda_{SO} > \lambda_R$, should still be achievable if accidental doping in the substrate and
cap layer can be limited.  When $\lambda_{SO}$ is smaller than $\lambda_R$, the energy gap closes and
graphene becomes a zero gap semiconductor with quadratic dispersion\cite{kane2005a}.

Our estimates suggest that the quantum spin Hall effect in graphene should
occur only below $\sim 0.01^\circ{\rm K}$, a temperature that is still accessible experimentally
but not as convenient as $\sim 1^\circ{\rm K}$.  In addition, it seems likely that disorder will
dominate over the spin-orbit couplings in current samples, so further progress in
increasing the mobility of graphene sheets may also be necessary before the
quantum spin Hall effect can be realized experimentally.  We emphasize, however\cite{sinitsyn2006}, that
the spin Hall effect survives, albeit with a reduced magnitude, even when the spin-orbit gap
is closed by disorder.

In summary, we have derived analytic expressions for the intrinsic and Rashba spin-orbit interaction
coupling constants that appear in the low-energy Hamiltonian of a graphene sheet under a perpendicular external electric field.
The Rashba interaction parameter is first order in the atomic carbon spin-orbit coupling strength $\xi$
and the perpendicular external electric field $E$, whereas the intrinsic spin-orbit interaction is second order in $\xi$
and independent of $E$. The estimated energy gap for $E=0$ is of the order of 0.01$^\circ {\rm K}$
and agrees with realistic {\it ab initio} electronic structure calculations.

\textit{Note added in proof.} Recently, we became aware of two
other articles which address spin-orbit interactions in
graphene and reach broadly similar conclusions\cite{note}.

\acknowledgements

This work was supported
by the Welch Foundation (Houston, TX) under Grant no. F-1473 and F-0934, by the Texas Advanced Computing Center (TACC), University of Texas at Austin, by Seagate Corporation, and by the Department of Energy under grant No. DE-FG03-96ER45598.
The authors gratefully acknowledge helpful conversations with A.K. Geim, C.L. Kane, and P. Kim.

\end{document}